\documentclass[12pt]{iopart}
\usepackage{graphicx}
\usepackage{amssymb}
\usepackage{color}
\usepackage{cite}

\begin{document}

\title{ Entanglement driven phase transitions in spin-orbital models }

\author{ Wen-Long You$^{1,2}$, Andrzej M. Ole\'s$^{1,3}$ and Peter Horsch$^1$ }

\address{$^1$ Max-Planck-Institut f\"ur Festk\"orperforschung,\\
              Heisenbergstrasse 1, D-70569 Stuttgart, Germany }
\address{$^2$ College of Physics, Optoelectronics and Energy, Soochow University,\\
              Suzhou, Jiangsu 215006, People's Republic of China }
\address{$^3$ Marian Smoluchowski Institute of Physics, Jagiellonian University,\\
              prof. S. \L{}ojasiewicza 11, PL-30348 Krak\'ow, Poland }

\ead{a.m.oles@fkf.mpg.de}

\date{\today}

\begin{abstract}
To demonstrate the role played by the von Neumann entropy spectra
in quantum phase transitions we investigate the one-dimensional
anisotropic SU(2)$\otimes XXZ$ spin-orbital model with negative
exchange parameter.
In the case of classical Ising orbital interactions we discover
an unexpected novel phase with Majumdar-Ghosh-like spin-singlet
dimer correlations triggered by spin-orbital entanglement and
having $k=\pi/2$ orbital correlations, while all the other phases
are disentangled. For anisotropic $XXZ$ orbital interactions both
spin-orbital entanglement and spin-dimer correlations extend to
the antiferro-spin/alternating-orbital phase. This quantum phase
provides a unique example of two coupled order parameters which
change the character of the phase transition from first-order to
continuous. Hereby we have established the von Neumann entropy
spectral function as a valuable tool to identify the change of
ground state degeneracies and of the spin-orbital entanglement
of elementary excitations in quantum phase transitions.
\end{abstract}


\submitto{\NJP}

\maketitle

\section{Spin-orbital physics and von Neumann entropy spectra}
\label{sec:intro}

In the Mott-insulating limit of a transition metal oxide the
low-energy physics can be described by Kugel-Khomskii-type models
\cite{Kug73}, where both spin and orbital degrees of freedom undergo
joint quantum fluctuations and novel types of spin-orbital order
\cite{Brz15} or disorder \cite{Cor12} may emerge.
Following the microscopic derivation from the multiorbital Hubbard
model, the generic structure of spin-orbital superexchange takes
the form of a generalized Heisenberg model \cite{Tok00,Ole05},
\begin{equation}
H=\sum_{\langle ij \rangle\parallel\gamma}\!\left\{
  J^{(\gamma)}_{ij}(\vec{T}_i,\vec{T}_j) \vec{S}_i\!\cdot\!\vec{S}_j
+ K^{(\gamma)}_{ij}(\vec{T}_i,\vec{T}_j)\right\},
\end{equation}
as indeed found not only for the simplest systems with $S=1/2$ spins:
KCuF$_3$ \cite{Kug73}, the RTiO$_3$ perovskites \cite{Kha00}, LiNiO$_2$
and NaNiO$_2$ \cite{Rei05}, Sr$_2$CuO$_3$ \cite{Woh13}, or alkali
RO$_2$ hyperoxides \cite{Sol08}, but also for larger spins as e.g. for
$S=2$ in LaMnO$_3$ \cite{Fei99}. In such models the parameters that
determine the spin-$S$ Heisenberg interactions stem from orbital
operators $J^{(\gamma)}_{ij}$ and $K^{(\gamma)}_{ij}$ --- they depend
on the bond direction and are controlled by the orbital degree of
freedom which is described by pseudospin operators $\{\vec{T}_i\}$.
That is, these parameters are not necessarily fixed by rigid orbital
order \cite{Zho08,Cor12}, but quantum fluctuations of orbital
occupation \cite{Kha05,Hor08} may strongly influence the form of the
orbital operators, particularly in states with spin-orbital
entanglement (SOE) \cite{Ole06,Ole12}. As a consequence, amplitudes
and even the signs of the effective exchange can fluctuate in time.
Such entangled spin-orbital degrees of freedom can form new states of
matter, as for instance the orbital-Peierls state observed at finite
temperature in YVO$_3$ \cite{Ulr03,Sir08}.
Another example are the collective spin and orbital excitations in a
one-dimensional (1D) spin-orbital chain under a crystal field which
can be universally described by fractionalized fermions \cite{Woh15}.
It is challenging to ask which measure of SOE would be the most
appropriate one to investigate quantum phase transitions in such systems.

The subject is rather general and it has become clear that
entanglement and other concepts from quantum information provide
a useful perspective for the understanding of electronic matter
\cite{Ami08,Zho06,Byc12,Hel13,Vee14}. Other examples of entangled
systems are:
topologically nontrivial states \cite{Hasan},
relativistic Mott insulators with $5d$ ions \cite{Jac09},
ultracold alkaline-earth atoms \cite{Gor10}, and
skyrmion lattices in the chiral metal MnSi~\cite{Muh09}.

One well-known characterization of a quantum system is the entanglement
entropy~(EE) determined by bipartitioning a system into $A$ and $B$
subsystems. This subdivision can refer for example to space
\cite{Ami08,Lun12}, momentum \cite{Tho10,Lun12}, or different degrees
of freedom such as spin and orbital \cite{You12}. A standard measure is
the von Neumann entropy (vNE),
${\cal S}_{\rm vN}^0\equiv -\textrm{Tr}_A\{\rho_A^{0}\log_2\rho_A^{0}\}$,
for the ground state $|\Psi_0\rangle$ which is obtained by integrating
the density matrix,
$\rho^{0}_A=\textrm{Tr}_B|\Psi_0\rangle\langle\Psi_0|$, over subsystem
$B$. Another important measure is the entanglement spectrum (ES)
introduced by Li and Haldane \cite{Li08}, which has been explored for
gapped 1D spin systems \cite{Pol10}, quantum Heisenberg ladders
\cite{Poi10}, topological insulators \cite{Fid10},
bilayers and spin-orbital systems \cite{Lun12}.
The ES is a {\it property of the ground state} and basically represents
the eigenvalues $p_i$ of the reduced density matrix $\rho^{0}_A$
obtained by bipartitioning of the system. Interestingly a
correspondence of the ES and the tower of excitations relevant for
SU(2) symmetry breaking has been pointed out recently \cite{Kol13}. It
was also noted that the ES can exhibit singular changes, although the
system remains in the same phase \cite{Lun14}. This suggests that the
ES has less universal character than initially assumed \cite{Cha14}.

In this paper we explore a different entanglement measure, namely the
vNE spectrum which monitors the vNE of ground and
\textit{excited states} of the system, for instance of a spin-orbital
system as defined in equation (1). In this case we consider the
entanglement obtained from the bipartitioning into spin and orbital
degrees of freedom in the entire system
\cite{You12}. Here the vNE is obtained from the density matrix,
$\rho^{(n)}_s=\textrm{Tr}_o|\Psi_n\rangle\langle\Psi_n|$,
by taking the trace over the orbital degrees of freedom (Tr$_o$)
for each eigenstate $|\Psi_n\rangle$.
We show below that the vNE spectrum,
\begin{eqnarray}
\label{spectra}
{\cal S}_{\rm vN}(\omega)=-\sum_n\textrm{Tr}_s\{\rho_s^{(n)}
\log_2\rho_s^{(n)}\}\delta\left\{\omega-\omega_n \right\},
\label{S_vN}
\end{eqnarray}
reflects the changes of SOE entropy for the different states at phase
transitions. The excitation energies, $\omega_n=E_n-E_0$, of
eigenstates $|\Psi_n\rangle$ are measured with respect to the ground
state energy $E_0$. It has already been shown that the vNE spectra
uncover a surprisingly large variation of entanglement within
elementary excitations \cite{You12}.
Also certain spectral functions have been proposed, that
can be determined by resonant inelastic x-ray scattering \cite{Ament},
and provide a measure of the vNE spectral function. Here we generalize
this function to arbitrary excitations $|\Psi_n\rangle$, i.e.,
\textit{beyond} elementary excitations which refer to a particular
ground state. We demonstrate that focusing on general
excited states opens up a new perspective that sheds light on quantum
phase transitions and the entanglement in spin-orbital systems.

The paper is organized as follows: In \sref{sec:som} we introduce the
1D spin-orbital model with ferromagnetic exchange, and in \sref{sec:phd}
we present its phase diagrams for the Ising limit of orbital
interactions and for the anisotropic SU(2)$\otimes XXZ$ model with
enhanced Ising component. SOE is analyzed in
\sref{sec:soe} using both the spin-orbital correlation function and
the entanglement entropy and we show that these two measures are
equivalent. In \sref{sec:es} we present the entanglement spectra and
discuss their relation to the quantum phase transitions. The main
conclusions and summary are given in \sref{sec:summa}.
The distance dependence of spin correlations in the antiferromagnetic
phase is explored in the Appendix.

\section{Ferromagnetic SU(2)$\otimes XXZ$ spin-orbital model}
\label{sec:som}

The motivation for our theoretical discussion of spin-orbital physics
comes from $t_{2g}$ electron systems in which orbital quantum
fluctuations are enhanced by an intrinsic reduction of the
dimensionality of the electronic structure \cite{Kha05}.
Examples of strongly entangled quasi-1D $t_{2g}$ spin-orbital systems
due to dimensional reduction are well known and we mention here just
 LaTiO$_3$ \cite{Kha00}, LaVO$_3$ and YVO$_3$ \cite{Hor08}, where the
 latter two involve $\{yz,zx\}$ orbitals along the $c$ cubic axis;
as well as $p_x$ and $p_y$ orbital systems in 1D fermionic optical
lattices \cite{Zhao,Sun12,Zho15}. This motivates us to consider the 1D
spin-orbital model for $S=1/2$ spins and $T=1/2$ orbitals with
anisotropic $XXZ$ interaction,
i.e., with reduced quantum fluctuation part in orbital interactions.
The $\vert+\rangle$ and $\vert-\rangle$ orbital states are a local
basis at each site and play a role of $yz$ and $zx$ states in $t_{2g}$
systems,
\begin{eqnarray}
\label{som}
{\cal H}(x,y,\Delta)&=&-J\sum_{j=1}^L
\left(\vec{S}_{j}\!\cdot\!\vec{S}_{j+1}+x\right)\!
\left([\vec{T}_{j}\!\cdot\!\vec{T}_{j+1}]_\Delta+y\right), \\
\label{delta}
[\vec{T}_{j}\cdot\vec{T}_{j+1}]_\Delta &\equiv&
\Delta\left(T_{j}^x T_{j+1}^x + T_{j}^y T_{j+1}^y\right)
+ T_{j}^z T_{j+1}^z,
\end{eqnarray}
where $J>0$ and we use periodic boundary conditions for a ring of
$L$ sites, i.e., $L+1\equiv 1$. The parameters of this model are
$\{x,y\}$ and $\Delta$. At~$x=y=1/4$ and $\Delta=1$ the model has SU(4)
symmetry. Hund's exchange coupling does not only modify $x$ and $y$ but
also leads to the $XXZ$ anisotropy ($\Delta<1$), a typical feature of
the orbital sector in real materials \cite{Ole05,Kha05}. The
antiferromagnetic model ($J=-1$) is Bethe-Ansatz integrable at the
SU(4) symmetric point \cite{Li99} and its phase diagram is well
established by numerical studies \cite{Ori00,Yam00}. It includes two
phases with dimer correlations \cite{Li05} which arise near the SU(4)
point. Some of its ground states could be even determined exactly at
selected $(x,y,\Delta)$ points \cite{Kol98,Kol01,Mar00,Kumar,Brz14}.

Here we are interested in the complementary and less explored model
with negative (ferromagnetic) coupling ($J=1$), possibly realized in
multi-well optical lattices \cite{Bel14}, which has been studied so
far only for SU(2) orbital interaction ($\Delta=1$) \cite{You12}. This
model is physically distinct from the antiferromagnetic ($J=-1$) model,
except for the Ising limit ($\Delta=0$) where the two models can be
mapped onto each other, but none was investigated so far.
The phase diagrams for $J=1$, see figure
\ref{phd}, determined using the fidelity susceptibility \cite{You07}
display a simple rule that the vNE (\ref{S_vN}) vanishes for exact
ground states of rings of length $L$ which can be written as products
of spin ($|\psi_s\rangle$) and orbital ($|\psi_o\rangle$) part,
$|\Psi_0\rangle=|\psi_s\rangle\otimes|\psi_o\rangle$.

\section{Phase transitions in the spin-orbital model}
\label{sec:phd}

\begin{figure}[t!]
\begin{center}
\includegraphics[width=13.4cm]{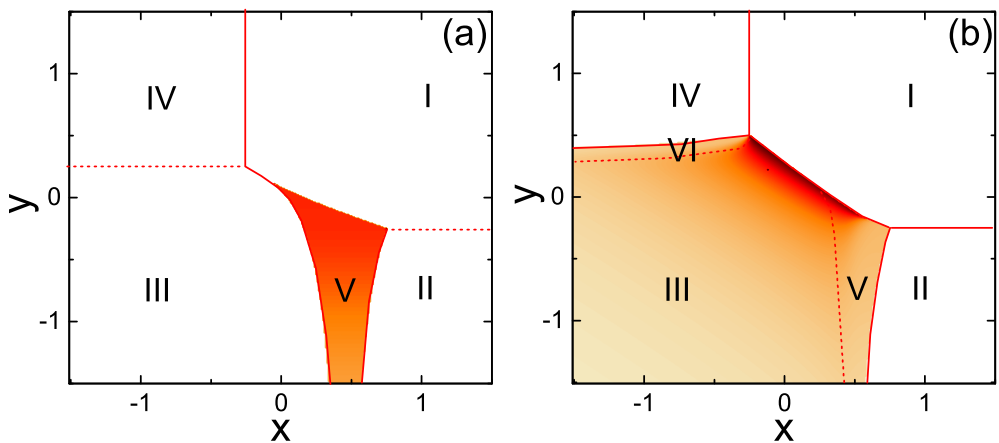}
\end{center}
\caption{Phase diagrams of the spin-orbital model
[equation (\ref{som})] obtained by two methods, fidelity susceptibility
or an exact diagonalization of an $L=8$ site model, for:
(a) $\Delta=0$, and
(b) $\Delta=0.5$.
The spin-orbital correlations in phases I-IV correspond to FS/FO,
AS/FO, AS/AO, FS/AO order (see text).
At $\Delta=0$ only the ground state of a novel phase V has finite EE,
${\cal S}^0_\textrm{vN}>0$ (shaded), whereas at $\Delta>0$ the EE
in phases III and VI is also finite.}
\label{phd}
\end{figure}

To understand the role played by the SOE in the 1D spin-orbital model
\eref{som} and \eref{delta} we consider the phase diagrams for
$\Delta=0$ and $\Delta=0.5$, see figure \ref{phd}. In the case
$\Delta=0$ all trivial combinations of ferro (F) and antiferro (A)
spin-orbital phases labeled I-IV have ${\cal S}^0_{\rm vN}=0$, i.e.,
spins and orbitals disentangle in all these ground states: FS/FO,
AS/FO, AS/AO, FS/AO. If both subsystems exhibit quantum fluctuations,
the ground state $|\Psi_0\rangle$ can no longer be written in the
product form. This occurs for the AS/AO phase III at $\Delta>0$.

\begin{figure}[t!]
\begin{center}
\includegraphics[width=10.5cm]{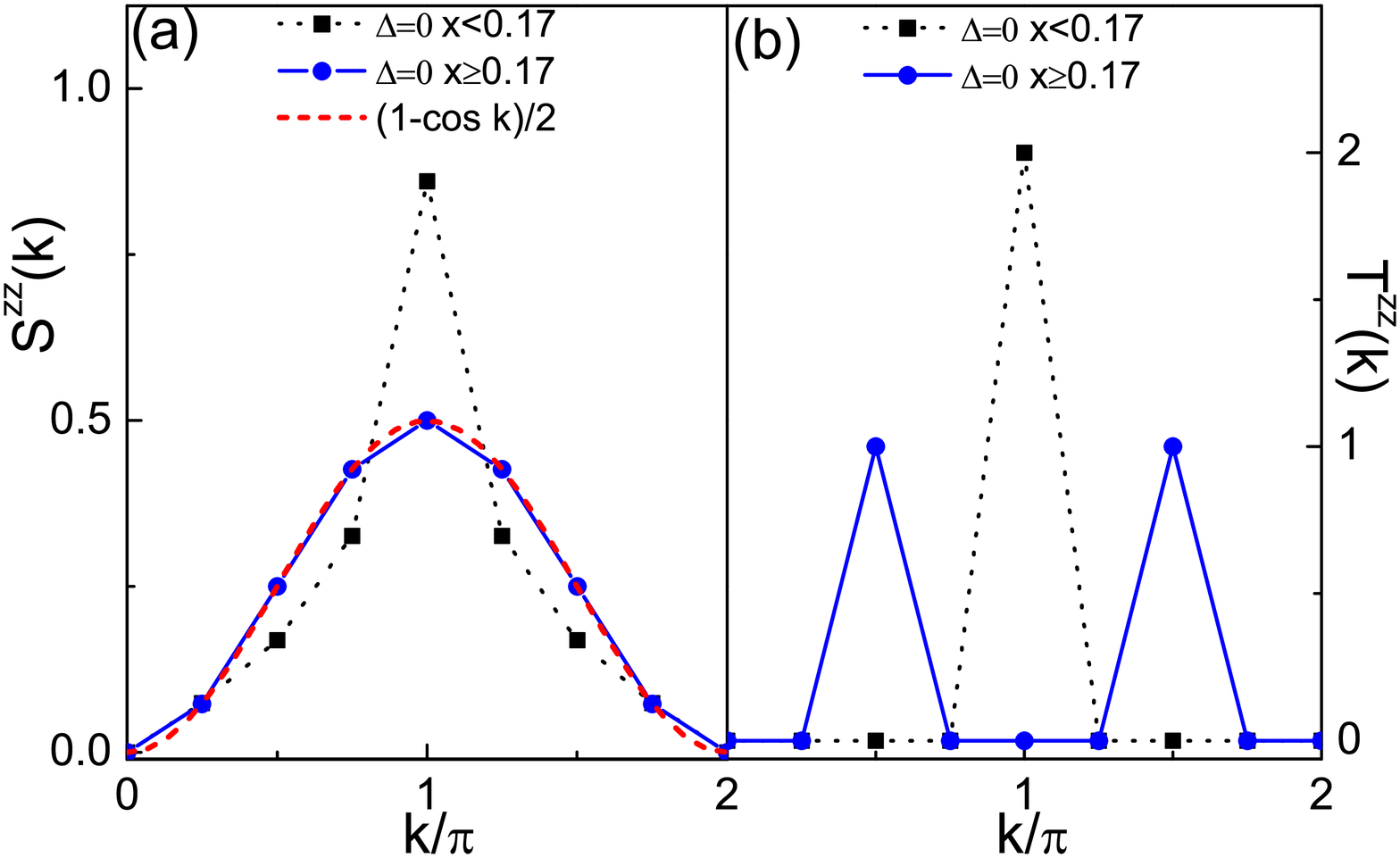}
\vskip -.4cm
\includegraphics[width=10.5cm]{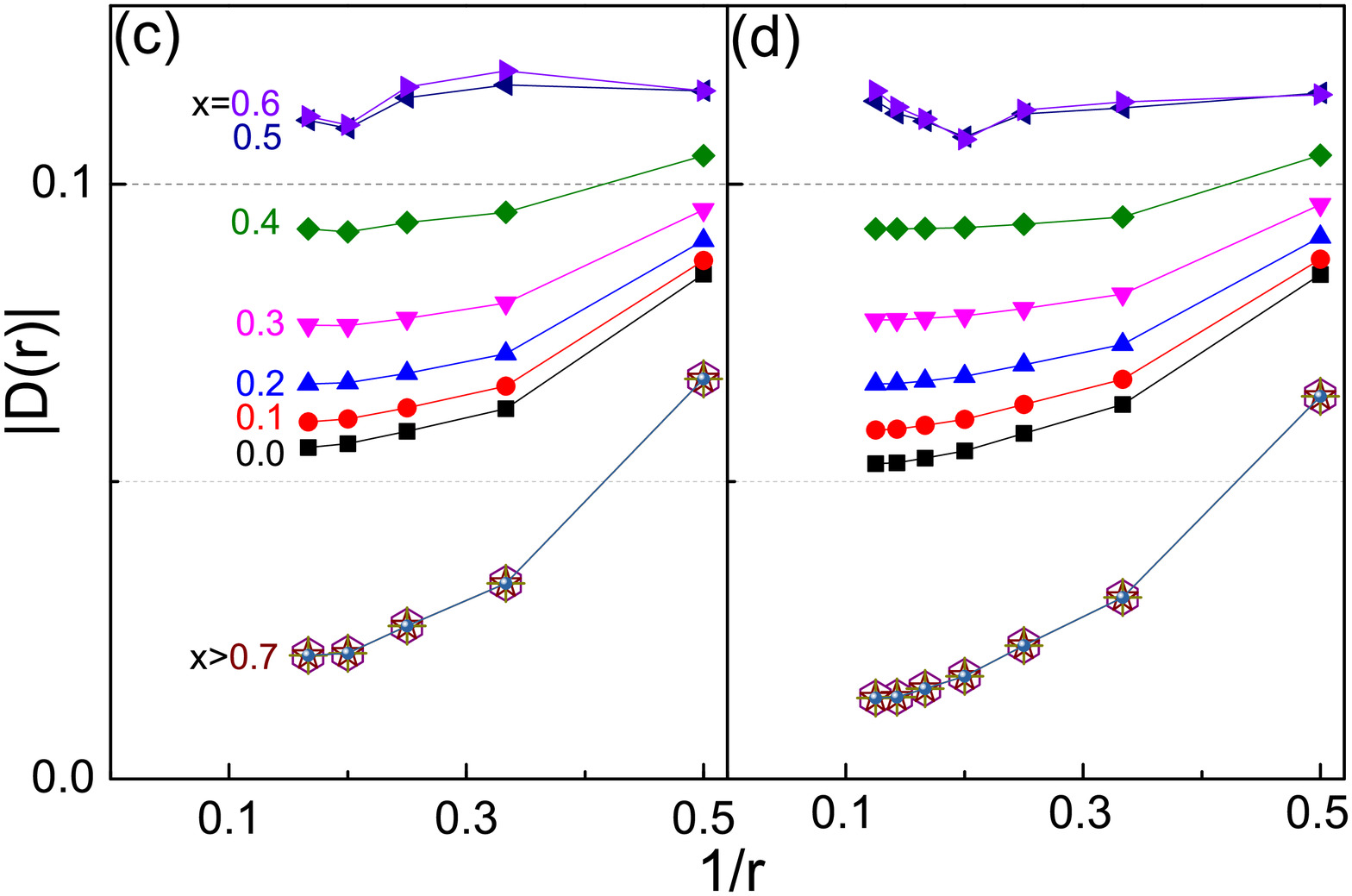}
\end{center}
\caption{
Top---
Spin [$S^{zz}(k)$] and orbital [$T^{zz}(k)$] structure factors
(\ref{sf}) for the 1D spin-orbital model (\ref{som}) of $L=8$
sites at $\Delta=0$ and $y=-1/4$:
(a) $S^{zz}(k)$ and
(b) $T^{zz}(k)$.
Bottom---
Spin dimer correlations $D(r)$ equation (\ref{D(r)}) found at $\Delta=0.5$
for decreasing $1/r$ for clusters of
(c) $L=12$ and
(d) $L=16$ sites.}
\label{Structurefactor}
\end{figure}

Most remarkable is the strongly entangled phase V at $\Delta=0$ and
$y<0$, see figure \ref{phd}(a). This phase occurs near $x\simeq
-\langle\vec{S}_j\!\cdot\!\vec{S}_{j+1}\rangle_{\rm AF}\equiv\ln 2-1/4$,
i.e., when the uniform antiferromagnetic spin correlations in phase III
are compensated by the parameter $x$, so that the energy associated with
Hamiltonian (\ref{som}) \textit{de facto disappears}. This triggers
state V with strong SOE (see below) as the only option for the system
to gain substantial energy in this parameter range by nonuniform
spin-orbital correlations. The analysis of phase V in terms of the
longitudinal equal-time spin (orbital) structure factors
\begin{eqnarray}
O^{zz}(k)&=&\frac{1}{L^2}\sum_{m,n=1}^{L}
e^{-ik(m-n)}\langle O_m^z O_n^z \rangle,
\label{sf}
\end{eqnarray}
where $O=S$ or $T$, reveals in figure \ref{Structurefactor}(a)
at $\Delta=0$ and $y=-1/4$  for the spin structure factor
$S^{zz}(k)\propto(1-\cos k)$.
This is a manifestation of nearest neighbour correlations, while
further neighbour spin correlations vanish and
moreover we find a quadrupling in the orbital sector, see figure
\ref{Structurefactor}(b). Thus the spin correlations indicate either
a short-range spin liquid or a translational invariant dimer state.

The {\it hidden spin-dimer order} \cite{Yon96} can be detected by the
four-spin correlator (we use periodic boundary conditions),
\begin{equation}
D(r)=\frac{1}{L}\sum_{j=1}^L\left[\left\langle
(\vec{S}_{j}\!\cdot\!\vec{S}_{j+1})
(\vec{S}_{j+r}\!\cdot\!\vec{S}_{j+1+r})\right\rangle
-\left\langle\vec{S}_{j}\!\cdot\!\vec{S}_{j+1}\right\rangle^2\right].
\label{D(r)}
\end{equation}
At $\Delta=0$ we find $|D(r)|$ with long-range dimer correlations in
phase V, but not in III. Phase III is a state with alternating ($k=\pi$)
spin (orbital) correlations in the range $x<0.17$ shown in figures
\ref{Structurefactor}(a,b).
Interestingly for $\Delta>0$ the dimer spin correlations $|D(r)|$ are
not only present in phase V but also appear in phase III. Moreover a
phase VI emerges, complementary to phase V, with interchanged role of
spins and orbitals, see figure \ref{phd}(b).
The order parameters for phase VI follow from the form of structure
factors which develop similar but complementary momentum dependence to
that for phase V seen in \fref{Structurefactor}(a), i.e., maxima at
$\pi/2$ for $S^{zz}(k)$ and at $\pi$ for $T^{zz}(k)$.
We remark that phases V and VI are unexpected and they were overlooked
before for the SU(2)$\otimes$SU(2) model at $\Delta=1$ \cite{You12}.
From the size dependence of $|D(r)|$ in figures \ref{Structurefactor}(c,d)
we conclude that the dimer correlations are long-ranged at $\Delta=0.5$
in phase V, but also in III, as seen from the data for $x\in[0.0,0.4)$,
where they coexist with the AS correlations.

These results suggest that the ground state V in \fref{phd}(a)
is formed by spin-singlet product states
\begin{eqnarray}
\vert\Phi_1^{D}\rangle&=&[1,2][3,4][5,6]\cdots[L-1,L], \nonumber \\
\vert\Phi_2^{D}\rangle&=&[2,3][4,5][6,7]\cdots[L,1],
\label{spins}
\end{eqnarray}
where $[l,l+1]=(\vert\!\uparrow\downarrow\rangle
-\vert\!\downarrow\uparrow\rangle)/\sqrt{2}$ denotes a spin singlet.
They are not coupled to orbital singlets on alternating bonds as it
happens for the AFantiferromagnetic SU(2)$\otimes$SU(2) spin-orbital
chain in a different parameter regime \cite{Kol98}, but to Ising
configurations in the orbital sector. The four-fold ($k=\pi/2$)
periodicity of orbital correlations is consistent with four orbital
states:
\begin{eqnarray}
\vert\Psi_1^{z}\rangle&=&\vert++--++\cdots --\rangle,  \nonumber \\
\vert\Psi_2^{z}\rangle&=&\vert-++--+\cdots +-\rangle,  \nonumber \\
\vert\Psi_3^{z}\rangle&=&\vert--++--\cdots ++\rangle,  \nonumber \\
\vert\Psi_4^{z}\rangle&=&\vert+--++-\cdots -+\rangle.
\label{orbis}
\end{eqnarray}
The decoupling of singlets is complete for $y=-1/4$ and $\Delta=0$,
where $(++)$ and $(--)$ bonds yield vanishing coupling in equation
(\ref{som}), and the phase boundaries of region V are $x^c_{\rm III,V}
=3/4+2\langle\vec{S}_j\!\cdot\!\vec{S}_{j+1}\rangle_{\rm AF}\simeq 0.136$
and $x^c_{\rm V,II}=3/4$ in the thermodynamic limit; moreover we find
perfect long-range order of spin singlets, i.e., $D(r)=(3/8)^2(-1)^r$.

\begin{figure}[b!]
\begin{center}
\includegraphics[width=11cm]{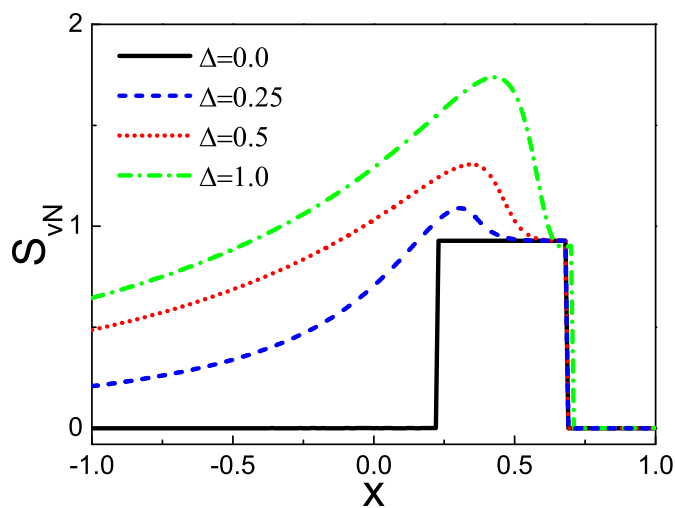}
\end{center}
\caption{Spin-orbital entanglement entropy ${\cal S}^0_\textrm{vN}$ in
the ground state of the spin-orbital model (\ref{som}) for the three
phases III, V and II as a function
of $x$ for various $\Delta$. Solid line for $\Delta=0$ stands for the
$k=0$ ground state in the limit of $\Delta\to 0$.
Parameters: $y=-0.5$ and $L=8$.}
\label{SvN-y=-0.5-x}
\end{figure}

The dimerized spin-singlet state at $\Delta=0$ has the same spin
structure as the Majumdar-Ghosh (MG) state \cite{Maj69}, however its
origin is different. While the MG state in a $J_1$-$J_2$ Heisenberg
chain results from frustration of antiferromagnetic exchange
(at $J_2=J_1/2$), here the spin singlets are induced by the SOE.
At $\Delta=0$ the only phase with finite SOE ${\cal S}_{\rm vN}^0=1$
is phase V, see figure \ref{SvN-y=-0.5-x}. In contrast, for $\Delta>0$
one finds finite EE also in phase III, when the original product
ground state changes into a more complex superposition of states and
joint spin-orbital fluctuations \cite{Ole06} appear. These
correlations control the SOE and give equivalent information to
${\cal S}^0_{\rm vN}$, see \sref{sec:soe}.
Furthermore, EE increases with $x$ towards phase V where it is further
amplified and exceeds ${\cal S}_{\rm vN}^0=1$.
The related softening of orbital order will be discussed below.
Interestingly we find a one-to-one correspondence of finite EE and
long-range order in the spin dimer correlations $|D(r)|$.

The superstructure of phase V emerges from the interplay of spin and
orbitals, where orbitals modulate the interaction of spins in equation
(\ref{som}), and \textit{vice versa}. It is important to distinguish
this from the Peierls effect, where the coupling to the lattice is
an essential mechanism. The orbital Peierls effect observed in
vanadates \cite{Ulr03,Sir08} or the orbital-selective Peierls
transition studied recently \cite{Str14} fall into the former category,
yet, as they involve orbital singlets ---
they are distinct from the case discussed here.

\section{Spin-orbital entanglement}
\label{sec:soe}

The description of spin-orbital entanglement in terms of the vNE
entropy, as discussed in \sref{sec:phd}, is a very convenient
measure of entanglement. But it is also a highly abstract measure.
To capture its meaning, one has to refer to mathematical intuition,
namely to the fact that any product state,
$|\Psi\rangle=|\psi_s\rangle\otimes|\psi_o\rangle$,
has zero vNE.
That is, an entangled state is a state that cannot be written as a
single product. A more physical measure are obviously spin-orbital
correlation functions relative to their mean-field value \cite{Ole06}.
Such correlation functions vanish for product states where mean-field
factorization of the relevant product is exact, i.e., spins and
orbitals are disentangled.

\begin{figure}[t!]
\begin{center}
\includegraphics[width=10.5cm]{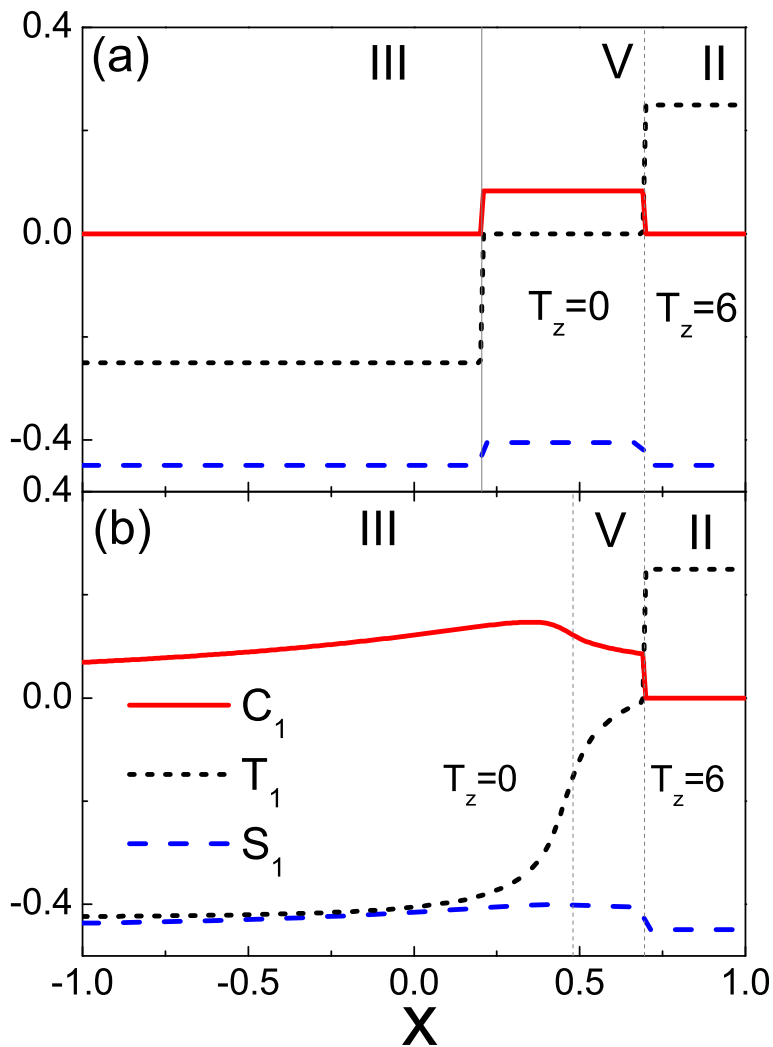}
\end{center}
\caption{Nearest neighbour
spin $S_{1}$ (\ref{sij}), orbital $T_{1}$ (\ref{tij}), and joint
spin-orbital $C_{1}$  (\ref{cij}) correlations as obtained for a
spin-orbital ring (\ref{som}) with $L=12$ sites and $y=-0.5$, as
functions of $x$ for:
(a) $\Delta=0$, and
(b) $\Delta=0.5$.
The III-V phase boundary (dotted vertical line) in (b) has been
determined by the maximum of the fidelity susceptibility
\cite{You07}.
}
\label{fig:cij}
\end{figure}

To detect spin-orbital entanglement in the ground state we evaluate
here the joint spin-orbital bond correlation function $C_1$ for the
SU(2)$\otimes XXZ$ model \eref{som}, defined as follows for a nearest
neighbour bond $\langle i,i+1\rangle$ in the ring of length $L$
\cite{Ole06},
\begin{equation}
\label{cij}
C_{1}\equiv\frac{1}{L}\sum_{i=1}^L
\Big\{\Big\langle({\vec S}_i\cdot{\vec S}_{i+1})
                 ({\vec T}_i\cdot{\vec T}_{i+1})\Big\rangle
      -\Big\langle{\vec S}_i\cdot{\vec S}_{i+1}\Big\rangle
       \Big\langle{\vec T}_i\cdot{\vec T}_{i+1}\Big\rangle\Big\}.
\end{equation}
The conventional intersite spin- and orbital correlation functions are:
\begin{eqnarray}
\label{sij}
S_{r}&\equiv&\frac{1}{L}\sum_{i=1}^L
\Big\langle{\vec S}_i\cdot {\vec S}_{i+r}\Big\rangle, \\
\label{tij}
T_{r}&\equiv& \frac{1}{L}\sum_{i=1}^L
\Big\langle{\vec T}_i\cdot {\vec T}_{i+r}\Big\rangle.
\end{eqnarray}
The above general expressions imply averaging over the exact
(translational invariant) ground state found from Lanczos
diagonalization of a ring.
While $S_{r}$ and $T_{r}$ correlations indicate the tendency towards
particular spin and orbital order, $C_{1}$ quantifies the spin-orbital
entanglement --- if $C_{1}\ne 0$ spin and orbital degrees of freedom
are entangled and the mean-field decoupling in equation (\ref{som})
cannot be applied as it generates uncontrollable errors.

Figures \ref{fig:cij}(a) and \ref{fig:cij}(b) show the nearest
neighbour correlation functions $S_1$, $T_1$ and $C_1$ at $y=-0.5$,
for $\Delta=0$ and $\Delta=0.5$, respectively, as functions of $x$. The
nearest neighbour spin correlation function $S_1$ is antiferromagnetic
(negative) in all phases III, V and II shown in figure \ref{fig:cij},
while (\textit{negative}) $T_1$ indicates AO correlations in phase III
and ferro-orbital (\textit{positive}) in phase II. Finite $\Delta=0.5$
triggers orbital fluctuations which lower $T_1$ below the classical
value of 0.25 found at $\Delta=0$. In the intermediate spin dimer phase
$T_1$ is negative for all $\Delta>0$, while it is zero for $\Delta=0$.

It is surprising that $C_1$ is positive in phase V at $\Delta=0$ in
spite of the classical Ising orbital interactions, see figure
\ref{fig:cij}(a). It is also positive in phases III and V at
$\Delta=0.5$ [see \fref{fig:cij}(b)]. Note that \textit{positive} $C_1$
is found in the present spin-orbital chain with $J>0$, while $C_1$ is
\textit{negative} when $J<0$ \cite{Li99}. In phase II $C_1$ vanishes in
the entire parameter range as then the ground state can be written as a
product. The same is true for phase III at $\Delta=0$. We emphasize
that the dependence of $C_1$ on $x$ is completely analogous to that of
the von Neumann entropy in \fref{SvN-y=-0.5-x}, which also displays
a broad maximum in the vicinity of the III-V phase transition at
$\Delta=0.5$, and a step-like structure in phase V at $\Delta=0$.
Thus we conclude here that the vNE yields a faithful measure of SOE in
the ground state that is qualitatively \textit{equivalent} to the more
direct entanglement measure via the spin-orbital correlation function
$C_1$ \cite{Ole06}.

\section{Entanglement spectra and quantum phase transitions}
\label{sec:es}

\begin{figure}[t!]
\begin{center}
\includegraphics[width=12.5cm]{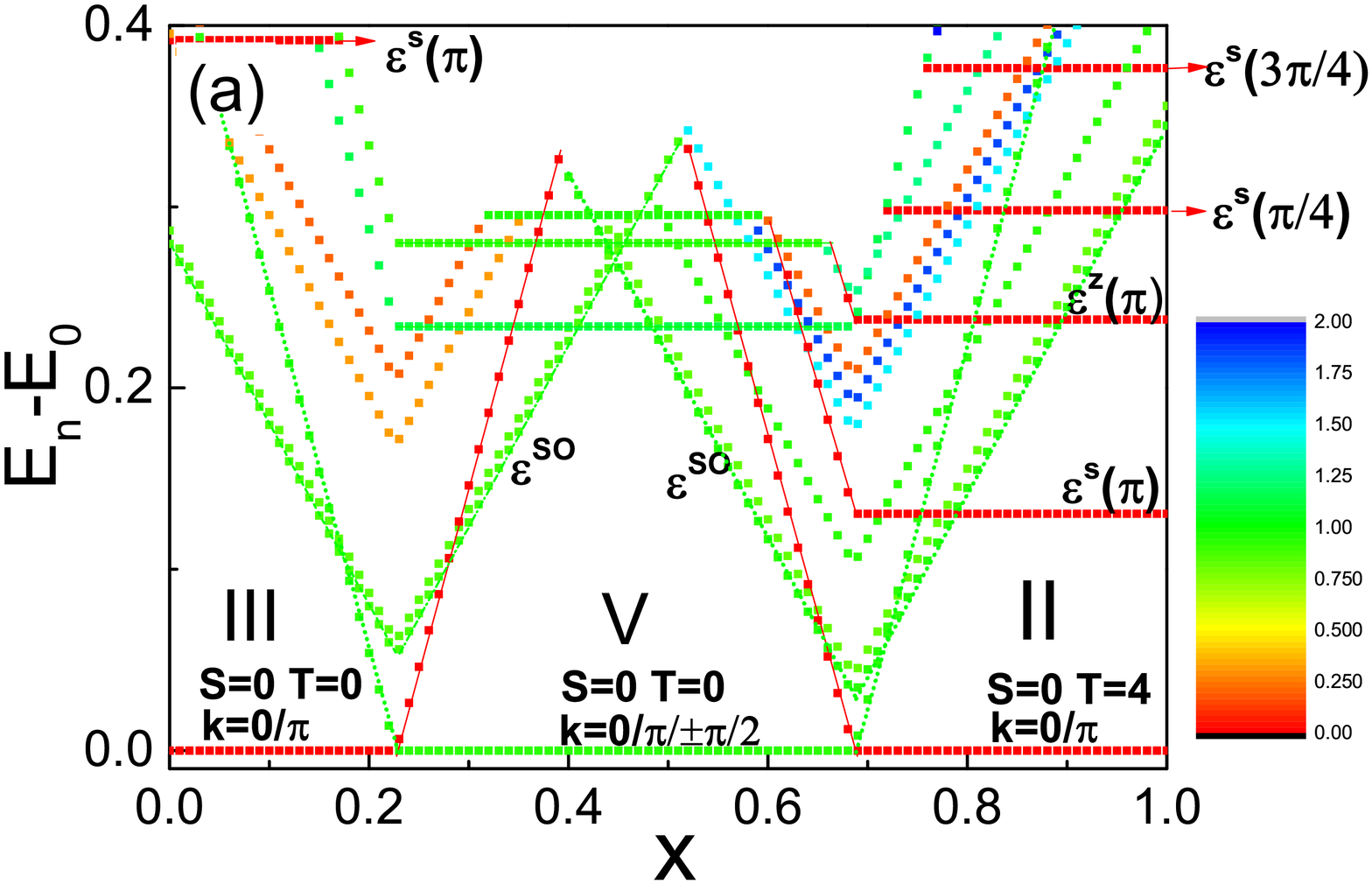}
\vskip -.2cm
\includegraphics[width=12.5cm]{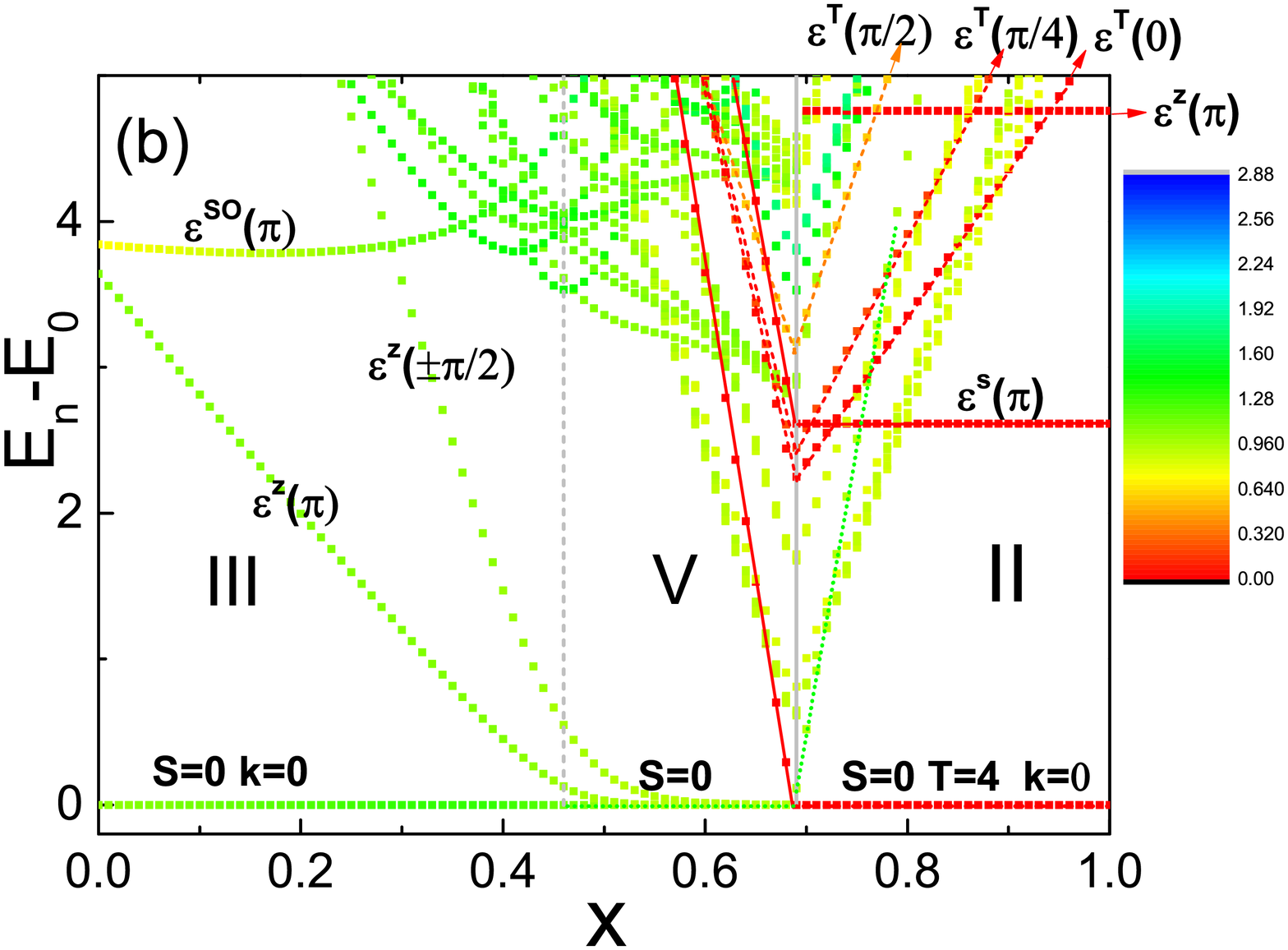}
\end{center}
\caption{
vNE-spectrum of lowest energies $E_n(x)$
(relative to the ground state energy $E_0(x)$) versus $x$ with
colors representing the size of the vNE of individual states. Data
for the three phases III, V and II is shown for $y=-0.5$, $L=8$ and:
(a) $\Delta=0$, and
(b) $\Delta=0.5$.
Here $\varepsilon^{\rm S}(k)$ [$\varepsilon^{\rm T}(k)$] denotes spin
(orbital) excitation, $\varepsilon^z(k)$ corresponds to an elementary
excitation having the same $S$ and $T$ as the ground state, and
$\varepsilon^{\rm SO}(k)$ stands for the spin excitation under
simultaneous flipping of orbitals.
}
\label{En-x-SvN-y=-0.25-Delta=0}
\end{figure}

Figure \ref{SvN-y=-0.5-x} stimulates the question about the origin
and the understanding  of the sudden or gradual EE changes at phase
transitions. This can be resolved by exploring the vNE spectral
function defined in equation (\ref{S_vN}) and shown in figures
\ref{En-x-SvN-y=-0.25-Delta=0}(a) and \ref{En-x-SvN-y=-0.25-Delta=0}(b)
for $\Delta=0$ and 0.5, where colors encode the vNE of states. The
excitation energies $\omega_n(x)=E_n(x)-E_0(x)$ are plotted here as
function of the parameter $x$. Only the lowest excitations are shown
that are relevant for the phase transitions and the low-temperature
physics. They include:
($i$) the elementary excitations of the respective ground state, and
($ii$) the many-body excited states that are relevant for the phase
transition(s) and may become ground states or elementary excitations
in neighbouring phases when the parameter $x$ is varied.

The AS/FO ground state of phase II in figure
\ref{En-x-SvN-y=-0.25-Delta=0}(a) obtained for a ring of $L=8$ sites is
an AS singlet ($S=0$) with a maximal orbital quantum number, $T=L/2=4$,
and a twofold ($k=0,\pi$) degeneracy at $\Delta=0$. The spin excitation
spectrum appears as horizontal (red) lines and consists of gapless
triplet $S=1$ excitations. The low-lying excitations
of the Bethe-Ansatz-solvable antiferromagnetic Heisenberg chain form
a two-spinon ($s$-$\bar{s}$) continuum, whose lower bound is given by
$\varepsilon(k)=\pi\vert\sin k\vert/2$ in the thermodynamic limit
\cite{Clo62}. For the $L=8$ ring the spectrum is discrete with a
$\Delta k=\pi/4$ spacing, and it is known that the energy of triplet
excitations $\varepsilon^S(\pi)$ will scale to zero as $1/L$
\cite{Hor88,Aff93,Kom94}. Red lines in phase II with finite slope are
orbital excitations. The $x$-dependence is due to the spin part of
$\cal H$ \eref{som} which determines both the orbital energy scale and
the dispersion,
$J_T\equiv(x+\langle\vec{S}_j\cdot\vec{S}_{j+1}\rangle_{\rm AF})
(1-\Delta\cos k)$.
This energy changes with $x$ and at finite $\Delta$ also with momentum
$k$, see figure~\ref{En-x-SvN-y=-0.25-Delta=0}(b). While the orbitons
are gapped, the low-lying excitations are either magnons or
$x$-dependent spin-orbital excitations. It is remarkable that the latter
are entangled in general, although the ground state II is disentangled.

\begin{figure}[t!]
\begin{center}
\includegraphics[width=10.5cm]{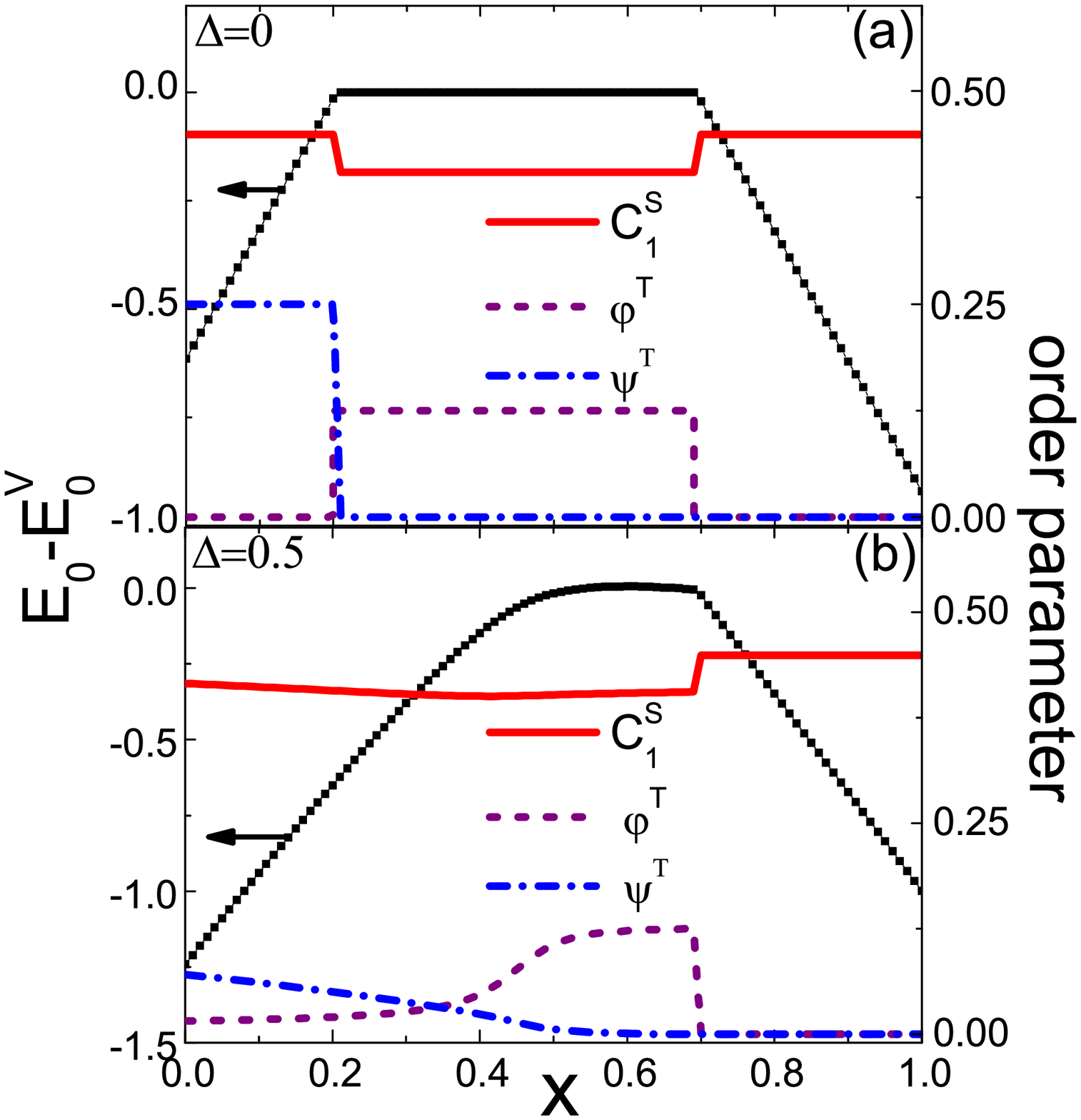}
\end{center}
\caption{Ground state energy relative to phase V,
$E_0(x)-E^V_0(x)$ (dots), orbital order parameters (dashed),
$\psi^T=[T^{\rm zz}(\pi)]^{\frac{1}{2}}$,
$\varphi^T=[T^{\rm zz}(\pi/2)]^{\frac{1}{2}}$,
and bond spin correlations
$|S_{1}|=|\langle \vec{S}_{1}\!\cdot\!\vec{S}_{2}\rangle|$ (solid line),
for phases III, V and II (from left to right), for:
(a) $\Delta=0.0$ and
(b) $\Delta=0.5$.
Parameters: $y=-0.5$ and $L=12$.
}
\label{OP-x-y=-0.25-Delta=0.1.eps}
\end{figure}

With decreasing $x$ a first-order phase transition from II to V occurs
by level crossing of disentangled (red) and entangled (green) ground
states. The spin-singlet ($S=0$) ground state of phase V has degeneracy
4 at $\Delta=0$, and its components are labeled by the momenta
$k=0,\pm\pi/2,\pi$. This is reflected by finite $\varphi^T$ order
parameter in figure \ref{OP-x-y=-0.25-Delta=0.1.eps}(a).
Note that at $\Delta>0$ this four-fold ground state degeneracy is lifted.
In the spin-dimer phase a gap opens in the spectrum of elementary spin
excitations \cite{Spr86,Uhrig}. The one-magnon triplet gap
$\Delta_S(\delta)\propto\delta^{3/4}$ depends on $y$ via the
dimerization parameter $\delta\equiv 1/\vert 4y\vert$. In phase V even
the pure magnetic excitations are entangled [see horizontal green
lines in figure \ref{En-x-SvN-y=-0.25-Delta=0}(a)]. The lowest
excitations in the vicinity of the phase transitions have orbital
character. From finite EE in figure \ref{En-x-SvN-y=-0.25-Delta=0}(a) one
recognizes that these states are inseparable spin-orbital excitations.

The phase transition from the dimer phase V to the AS phase III ($S=0$)
appears singular in the sense that it is first order at $\Delta=0$ and
continuous otherwise [figures \ref{En-x-SvN-y=-0.25-Delta=0}(a,b)].
To locate the center of the continuous phase transition between phases
III and V at $\Delta>0$, we have selected the peak of the first
derivative of the entanglement entropy, see \fref{SvN-y=-0.5-x}.
Yet also the peaks in the derivatives of the fidelity susceptibility,
the orbital correlation function $T_1$ [see \fref{fig:cij}(b)] and the
orbital order parameters $\psi^T$ and $\varphi^T$ in
\fref{OP-x-y=-0.25-Delta=0.1.eps}(b) may be used. Finally, we note
that the scaling of entanglement with system size has quite different
behaviour in phases III and V, indicating that a phase transition
separates them.

Furthermore, the peculiar feature of the AS/AO phase III manifests
itself in a twofold degeneracy and zero SOE at $\Delta=0$ in contrast
to the nondegenerate ground state and finite SOE at finite $\Delta$.
The entanglement has two sources, namely:
($i$) the interplay of quantum fluctuations in the spin and orbital
sectors and
($ii$) the dimerization order which coexists
with antiferromagnetic spin correlations in phase III at finite $\Delta$.
The latter is the origin of the nondegenerate ground state as it yields
a coupling to the $\varepsilon^{\rm SO}(\pi)$ excitation (nearly
horizontal in $x$), and the emergence of the spin-dimer correlations
$D(r)$ leads to a faster decay of the spin correlations in phase III
than in the 1D antiferromagnetic Heisenberg chain, see the Appendix.
The orbital order parameters $\psi^T$ and $\varphi^T$ compete in
phases III and V, see figure \ref{OP-x-y=-0.25-Delta=0.1.eps}(b), near
the phase boundary in figure \ref{phd}(b). This also explains why the
transition from phase V to III is smooth at finite $\Delta$ in terms of
both the vNE (figure~\ref{SvN-y=-0.5-x}) and the nearest neighbour spin
correlations $|S_1|$.

\section{Conclusions and summary}
\label{sec:summa}

Summarizing, we have studied the quantum phases and the spin-orbital
entanglement of the 1D ferromagnetic SU(2)$\otimes XXZ$ model by means
of the Lanczos method. We have discovered a previously unknown
translational invariant phase V with long-range spin singlet order and
four-fold periodicity in the orbital sector. Its mechanism is distinct
from the dimer phases found in the 1D antiferromagnetic spin-orbital
model near the SU(4) symmetric point \cite{Li05}. Both III-V and II-V
phase transitions arise from the spin-orbital entanglement in the case
of Ising orbital interactions. When the orbital interactions change
from Ising to anisotropic $XXZ$-type, the entanglement develops in
phase III, where antiferromagnetic spin correlations and long-range
spin dimer order coexist, changing the quantum phase transition from
first-order to continuous. Furthermore in the regime of finite orbital
fluctuations ($\Delta>0$) another phase VI emerges, which is
complementary in many aspects to phase V, but with the important
difference that phase VI disappears in the limit $\Delta=0$.

We have shown that the von Neumann entropy spectral function
${\cal S}_{\rm vN}(\omega)$ \eref{S_vN} is a valuable tool that
captures the spin-orbital entanglement SOE of excitations and explains
the origin of the entanglement entropy change at a phase transition.
From the perspective of spin-orbital entanglement we encounter
($i$) first-order transitions between disentangled (II) and entangled
(V) phases,
($ii$)~a continuous transition involving two competing order parameters
between two entangled phases, III and V, and
($iii$) trivial first-order transitions between two disentangled phases.
Case ($ii$) goes beyond the commonly accepted paradigm of a single
order parameter to characterize a quantum phase.

Moreover, we have presented two simple measures of entanglement in the
ground state and shown that they are basically equivalent --- the direct
measure via the (quartic) spin-orbital bond correlation function $C_1$
\eref{cij} and the von Neumann entropy ${\cal S}^0_{\rm vN}$. The latter
is defined by separating globally spin from orbital degrees of freedom
in the ground state.

\ack

We thank Bruce Normand and Krzysztof Wohlfeld for insightful
discussions. W-L You acknowledges support by the Natural
Science Foundation of Jiangsu Province of China under
Grant No. BK20141190 and the NSFC under Grant No. 11474211.
A M Ole\'s kindly acknowledges support by Narodowe Centrum Nauki
(NCN, National Science Center) under Project No. 2012/04/A/ST3/00331.

\section*{Appendix: Distance dependence of the antiferromagnetic spin
correlations}
\label{sec:spin}

Here we explore in more detail the competition of the antiferromagnetic
(AF) spin correlations of the spin-orbital chain in the AS/AO phase III
and the Majumdar-Ghosh like spin-singlet dimer correlations that coexist
at finite $\Delta$, as we found in our work. For $\Delta=0$ the spin
correlations in phase III are those of an AF Heisenberg spin chain,
\begin{eqnarray}
\left\langle {\vec S}_i\cdot {\vec S}_{i+r}\right\rangle\sim
(-1)^{r}\frac{\sqrt{\ln\vert r\vert}}{\vert r\vert},
\end{eqnarray}
which reveal the typical $1/r$-power law decay combined with
logarithmic corrections that were first predicted by conformal field
theory \cite{Aff89,Luk98} as well as by renormalization group methods
\cite{Gia89}, and subsequently confirmed \cite{Hal95} by
numerical density matrix method \cite{Sch05}.

\begin{figure}[t!]
\begin{center}
\includegraphics[width=11cm]{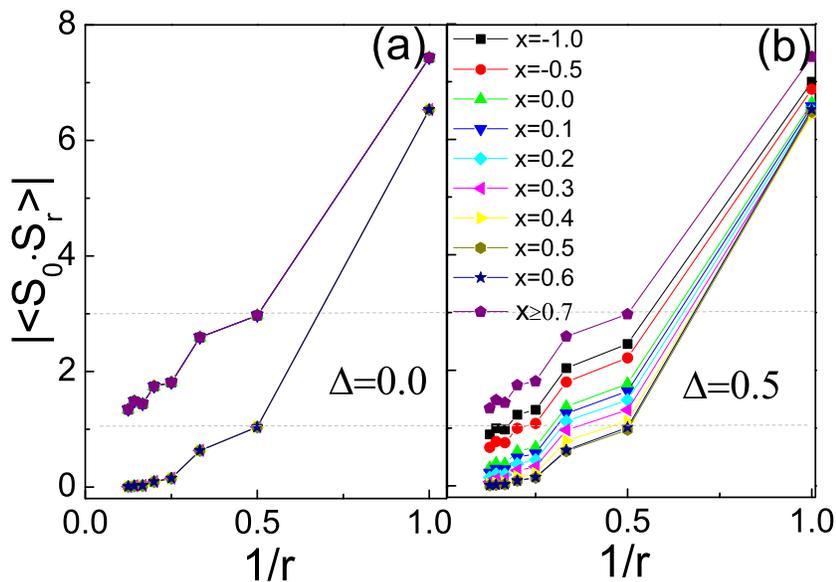}
\end{center}
\caption{
Modulus of spin correlations $S_r$ equation (\ref{sij}) versus the
inverse distance $1/r$ as obtained for a spin-orbital ring (\ref{som})
with $L=16$ sites, for:
(a) $\Delta=0$, and
(b) $\Delta=0.5$.
Parameter: $y=-0.5$.}
\label{fig:s0sr}
\end{figure}

In figure \ref{fig:s0sr}(a) we present our numerical data for the
spin-correlation function $S_r$ equation (\ref{sij}) (i.e., for
translational invariant ground states) for several values of $x$, and
for $\Delta=0$ and $y=-0.5$. In the $\Delta=0$ case there are only two
distinct types of behaviour of $S_r$, namely exponential decay in phase
V and the power law decay of the 1D quantum N\'eel spin liquid state,
which are the same in phases II and III.

\begin{figure}[t!]
\begin{center}
\includegraphics[width=11.5cm]{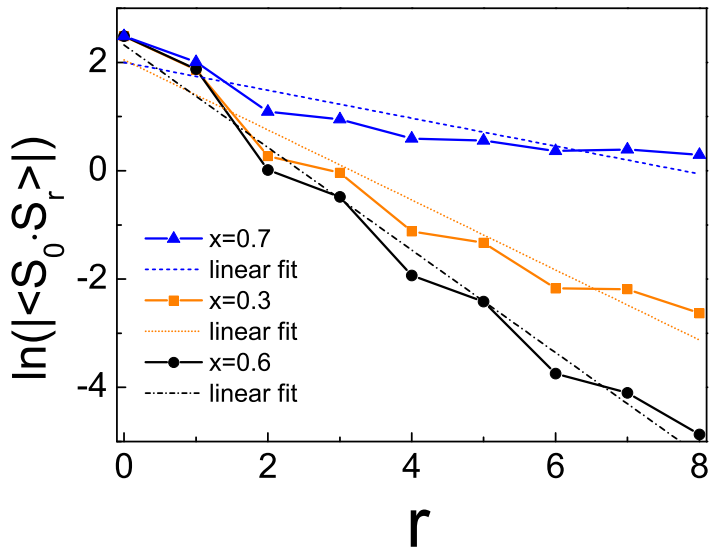}
\end{center}
\caption{
Logarithm of modulus of spin correlations $S_r$ equation (\ref{sij})
for increasing distance $r$ as obtained for the spin-orbital model
equation (\ref{som}) on a ring of $L=16$ sites for $\Delta=0.5$,
$y=-0.5$, and three values of $x$. Exponential decay of $S_{r}$
with increasing $r$ is obtained for phase V ($x=0.6$).}
\label{fig:logfit}
\end{figure}

Figure \ref{fig:s0sr}(b) displays $S_r$ at $\Delta=0.5$ for different
$x$-values. Here again the unperturbed AF correlations of the 1D N\'eel
spin-liquid state appear in phase II ($x\ge 0.7$). It is evident that
in phase III the AF spin correlations are strongly reduced, due to the
competition with the coexisting long-range ordered spin-singlet
correlations. The spin singlet order increases with $x$ in phase III,
and as a consequence we observe here that the decay of $S_r$ becomes
stronger as $x$ approaches the III/V transition.

In figure \ref{fig:logfit} we present a logarithmic plot which
highlights the different decays of $S_r$ for $\Delta=0.5$ in the three
different phases: III, V, and II. We have selected the values for
$x=0.3$, $0.6$ and $0.7$, respectively, for greater transparency.
The log-plot shows clearly the exponential decay of $S_r$ in phase V.
It also shows that the $L=16$ system reveals strong finite size effects
in phase II where $S_r$ has power law decay. Nevertheless it is clear
already from the $L=16$ data that the AF spin correlations in phase III
(here shown for $x=0.3$) are strongly suppressed and approach the
exponential decay of $S_r$ in phase V ($x=0.5$ and 0.6) when $x$
approaches the III-V phase boundary from the left.

Summarizing, we find that in phase III the AF spin correlations of the
1D N\'eel spin liquid state decay much more rapidly as the competing
spin-singlet order emerges. This effect is particularly strong near
the boundary of phase III to the spin-singlet dimer phase V. Whether
in the thermodynamic limit the correlations $S_r$ also decay
exponentially in phase III as in V cannot be decided here,
and this question is beyond the scope of the present work.

\end{document}